\documentclass[12pt]{article}
\newcommand{\half}{{\textstyle{\frac{1}{2}}}}
\newcommand{\be}{\begin{equation}}
\newcommand{\ee}{\end{equation}}
\newcommand{\bea}{\begin{eqnarray}}
\newcommand{\eea}{\end{eqnarray}}

\newcommand{\lra}{\Leftrightarrow}

\begin{document}
\newpage
\pagestyle{empty}
  
 \vspace{10mm}

\begin{center}

\vspace{10mm}

{\Large \textbf{ From quantum groups to genetic mutations }}

\vspace{10mm}

{\large     A. Sciarrino}

\vspace{10mm}

 \emph{Dipartimento di Scienze Fisiche, Universit{\`a} di Napoli
``Federico II''}

\emph{  I.N.F.N., Sezione di Napoli}

\emph{Complesso Universitario di Monte S. Angelo}

\emph{Via Cintia, I-80126 Napoli, Italy}

 \end{center}

\vspace{7mm}

\begin{abstract}
 In the framework of the crystal 
basis model of the genetic code, where each codon is assigned to an 
irreducible representation of  $U_{q \to 0}(sl(2) \oplus sl(2))$, 
single base mutation matrices
   are  introduced. The strength  of the mutation is assumed to
depend on the ``distance" between the codons. Preliminary general predictions of the model are
 compared with experimental data, with a satisfactory agreement.
  
\end{abstract}

\vfill
\begin{center}
Symmetries in Science XIII \\
Bregenz, July 20-24 2003
\end{center} 
\vfill

\rightline{DSF-28/03}
\rightline{  }
 
\vspace*{3mm}
\hrule
\vspace*{3mm}
\noindent
\emph{E-mail:}
  \texttt{sciarrino@na.infn.it}

\newpage
 \phantom{blankpage}
\newpage
\pagestyle{plain}
\setcounter{page}{1}
\baselineskip=16pt

\section{Introduction}

Among the numerous and important questions offered to the theoretical physicist 
by the sciences of life, the ones relative to the genetic code present a 
particular interest.  The DNA structure   and the mechanism
 of polypeptid fixation from codons   possess appealing aspects 
for the theorist and, indeed, the first proposal of genetic code may 
be ascribed to G. Gamow \cite{Gamow} in 1954, less than year after the discovery 
of   DNA by Watson and Crick. Let us briefly recall
 some essential features, see e.g. \cite{livre}. First  the DNA macromolecule is 
constituted by two linear chains of nucleotides in a double helix shape. 
There are four different nucleotides, characterised by their bases: adenine 
(A) and guanine (G)   (purines family),  cytosine (C) and thymine (T) 
  (pyrimidines family). Note also that an A (resp. T) base in one strand is 
connected with two hydrogen bonds to a T (resp. A) base in the other 
strand, while a C (resp. G) base is related to a G (resp. C) base with 
three hydrogen bonds. The genetic information is transmitted via the messenger
 ribonucleic acid or mRNA. During this 
operation, called transcription, the A, G, C, T bases in one strand of the DNA 
are associated respectively to the U, C, G, A bases, (U denoting the uracile 
base) of RNA. Then,  a triplet of nucleotides or 
codon will be related to an amino-acid. More precisely, a codon is defined 
as an ordered sequence of three nucleotides, e.g. AAG, AGA and GAA, and one 
enumerates in this way $4 \, \times \, 4 \, \times \, 4 = 64$ different 
codons.  In the universal eukariotic code (see Table 
\ref{table:rep}), 61 of such triplets  encode  the amino-acids, 
 while the three codons UAA, UAG and 
UGA, which are called non-sense or stop-codons,  play the role to 
stop the biosynthesis process. Indeed, the genetic code is the association 
between codons and amino-acids. But since one distinguishes only 20 
amino-acids \footnote{Alanine (Ala), Arginine (Arg), Asparagine (Asn), 
Aspartic acid (Asp), Cysteine (Cys), Glutamine (Gln), Glutamic acid (Glu), 
Glycine (Gly), Histidine (His), Isoleucine (Ile), Leucine (Leu), Lysine 
(Lys), Methionine (Met), Phenylalanine (Phe), Proline (Pro), Serine (Ser), 
Threonine (Thr), Tryptophane (Trp), Tyrosine (Tyr), Valine (Val).} related 
to the 61 codons, it follows that the genetic code is degenerated.   
 From Table \ref{table:rep}, one remarks the presence of 3 sextets, 
5 quadruplets, 1 triplet, 9 doublets and 2 singlets of codons, each multiplet 
corresponding to a specific amino-acid. Since its appearance on the 
earth life has been characterized by its continuous change.
 Spontaneous {\bf genetic mutations}, i.e. modifications of the DNA genomic
 sequences, play
a fundamental role in the evolution.  In the present paper I only deal 
with  point mutations, that is with  single 
base (single nucleotide) changes. More generally, mutations include changes
of more than one nucleotide, insertions and 
deletions of nucleotides, frame-shifts and inversions. The point mutations are usually 
modeled by 
stationary, homogeneous Markov process, which assume: 

\noindent 1) the nucleotide 
positions are stochastically independent one from another, which is clearly 
not true in functional sequences; 

\noindent2) the mutation is not
depending on the site and constant in time, which  ignores the existence 
of ``hot spots" for mutations  as well as the probable existence of 
evolutionary spurts; 

\noindent 3) the nucleotide frequencies  
are equilibrium frequencies. Moreover a common belief is that tha 
change 
of the 3rd nucleotide is more frequent than the change of the 1st 
nucleotide, the latter being more frequent than the change of the 
second one
  
In the following the labels $i,j$ run in the set analysed, e.g. 
$i,j \in \{C,T,G,A \}$ ($T$ being replaced by $U$ in RNA) for single 
nucleotides changes or $i,j$  run in a 20-dim set for the amino-acids 
substitution matrix or in a 64-dim set for for the codon substitution matrix.
The transition matrix ${\bf Q}$, where $Q_{ij} > 0$ ($i \neq j$)
represents the transition rate between the $j$ state and the $i$ state,
 in the choosen unit of ``time",  and it is normalised to
\be
 0 > Q_{ii} = 1 \, - \, \sum_{j \neq i} \, Q_{ij} 
\ee
The evolution matrix  ${\bf P}$, where $P_{ij}(t)$ gives the probability that 
the $j$ state at time $t = 0$,  will be replaced, at time $t$,  by the 
$i$ state, satisfies the differential equation  
\be
\frac{d P_{ij}(t)}{dt}  = \sum_{k} \,   P_{ik}(t) \,  Q_{kj} \;\; \lra \;\;
{\bf P}(t) ={\bf P}(0) \exp{{\bf Q} \, t} \;\;\;\; {\bf P}(0)  = {\bf 1} \label{eq:P}
\ee
In the Markov model, with discretized time $\tau$, we have
\be
{\bf P}((n + 1) \tau) ={\bf Q \, P}(n \tau)  
\ee
 The most simple  reversible model describing single nucleotide 
  changes
depends on 1 parameter and the most complex not reversible model depends on 
12 parameters \cite{ROM}. \footnote{For a review of the different Markov models 
with a large list of the original papers, 
see Cap. 3 of \cite{Li}.} 
These models consider the DNA
 sequences as set of nucleotides each nucleotide evolving independently of the 
 others; they are not able to make, a priori, any 
prediction on  the reversibility of a mutation and naturally predict that 
a  nucleotide change happens at the same rate independently of which codon 
it belongs to. The following shortcomings are particularly serious:  the 
Markov models are indeed unable to explain

\noindent i) the dependence of mutations  on the nature  of the neighbouring 
nucleotides \cite{Blake}. These features can of course be accounted 
introducing more new unknown parameters or new type of models, see 
\cite{ABH};

\noindent ii) the fact that mutations  occur more frequently between amino acids
with similar physico-chemical properties, which generally have similar 
 functional roles. Generally in the literature it is stated that the 
 nature of the 2nd nucleotide strongly determines the 
 physico-chemical propertie. In the seventies Konopolchenko and Rumer \cite{KR} 
 have remarked that amino acids with similar physico-chemical 
 properties can be described by assigning a suitable charge $Q$ to 
 the first dinucleotide (called ``root" by the authors) of the codon, 
 in particular ``strong roots" (``weak roots'), corresponding to multiplets of 
 codons of dimension $ 4$ ($\leq 3$), have $Q > 0$ ($Q < 0$). Note 
 that sextets appear as the sum of a quartet and of a doublet. 

The aim of this paper is to propose a  model in which the strength of the 
 mutation depends on a suitably defined {\bf distance} between codons.
 This model reduces to the Markov model if the distance dependence is 
 assumed constant, but it is able, in principle, to take into some account the 
points i)-ii). The first requirement to build such a model  
 is to identify codons as mathematical objects, in particular  
 as  vectors in a  suitable space. This will be done
in the framework of the {\it crystal basis model} of the genetic code
 \cite{FSS}. In this model the 4 nucleotides  are assigned to 
the (4-dim  fundamental) irreducible representation (irrep.) 
$(1/2, 1/2)$ of $U_{q \to 0}(sl(2) \oplus sl(2))$ 
with the following assignment for the values of  the third 
component of  $\vec{J}$ for the two $sl(2)$ which in the following will be 
denoted as  $sl_{H}(2) $   and $sl_{V}(2)$:
\be 
	\mbox{C} \equiv (+\half,+\half) \qquad \mbox{T/U} \equiv 
(-\half,+\half) 
	\qquad \mbox{G} \equiv (+\half,-\half) \qquad \mbox{A} \equiv 
	(-\half,-\half) \label{eq:gc1}
	\ee
	and the codons, triple of nucleotides, to the $3$-fold tensor product of 
 $(1/2, 1/2)$.  The assignment of the codons to the different irreps.  and 
the correspondence with the encoded  amino acid in  the eukaryotic  code  
 is provided in Table \ref{table:rep}.  Let us emphasize that the assignment of 
 the codons to the different irreps. is a 
  straightforward consequence of the assumed  labelling of the nucleotides 
  eq.(\ref{eq:gc1}) and of the Kashiwara's theorem on the tensor product of irreps. in 
the crystal basis  \cite{Kashi}.  In the following we call
{\bf nearest codon}  codons differing by only one nucleotide.  The effects of 
a single nucleotide mutation in the codons   are  
represented, neglecting the mutations into or from the three stop codons 
which are not detectable in the considered set of experimental data, 
by a $61 x 61$ (symmetric) matrix,
 whose elements, in first 
approximation, will be assumed vanishing if non connecting nearest codons. 
 
In ref. \cite{FSSp} it has been shown that amino acids with similar 
properties can grouped together looking to the content of the irrep. of the first 
dinucleotide (or ``root"), in particular to the values of the charge $Q$ and the 
third generator of $sl_{V}(2)$.  The charge $Q$ can be expressed as\footnote{Note that
 the numerical values of 
	eq.(\ref{eq:Q}) are slightly different from  those  of \cite{KR}.}    
	 \be
 Q = 4 \, J_{3,H} \, + \, C_{V}(J_{3,V} + 1) \, - \, 1  
 \label{eq:Q}
 \ee
 In that paper the analysis has been performed for 10 
physico-chemical properties: the  Chou-Fasman 
conformational parameters, which give
 a measure of the probability of
	the amino acid to form respectively a helix, a sheet and a turn;
	the Grantham polarity; the relative hydrophilicity;  
	the thermodynamic activation parameters at 298 K: $\Delta H$
(enthalpy, in kJ/mol), $\Delta G$ (free energy, in kJ/mol) and $\Delta S$
	(entropy, in J/mole/K); the dissociation constants at 298 K; the isoelectronic point, i.e. the $pH$ value at which
	no electrophoresis occurs. 
   The strength of the mutation inducing operator
is assumed to depend on the distance between the initial codon and the final 
codon, i.e. the codon appearing as  result of the mutation. In the literature
many attempts to define distance between codons exist based  on the 
similarity of their physico-chemical properties or of those of the encoded 
amino-acid. Sometimes the distance between amino acids is defined  by the 
strength of their  mutation. Here I follow a completely different approach as I define a 
 priori a distance and then I try to derive the strength of their 
 mutation.  
 
\section{The mutation matrix}

In order to be able to define the distance we make a correspondence between 
a codon and a point in $n$-dim. Euclidean space. For sake of 
simplicity, presently we assume a 
$1$-dim space.\footnote {The use of a 2-dim space, related to the   
 roots of the  two commuting $sl(2)$, may seem the most naturale 
choice.} The correspondence between  codons and   real numbers is 
realized through the eigenvalues of the following operator
\be
     \hat{X}  =  [  \, \alpha  \,   Q^{1}   \, - \, \beta \, 
  J_{3,V}^{1} (J_{3,V}^{1} - 1) \,
  +  \,  4  \gamma  \, (C_{H} + C_{V}) ] \; 2  \, (J_{3,H}  + \eta J_{3,V})  
\label{eq:oper} 
\ee
where  $\alpha, \beta, \gamma$ and $\eta$  are real positive parameters
 ($\eta > 1$ as mutations between pyrimidines and purines 
 (tranversions, $ \Delta J_{3,V} \neq 0$) occur less frequently than 
 mutations between pyrimidines or purines (transitions $ \Delta 
 J_{3,H} \neq 0$)); $ Q^{1}$ and  $J_{3,V}^{1} $ are, respectively,  
  the ``charge", given by eq.(\ref{eq:Q}), and   the third  generators
 of $sl_{V}(2)$ of the  first dinucleotide of the codon $XYZ$, that is  $XY$, 
 and $C_{H}, J_{3,H}$ (resp. $ C_{V}, J_{3,V}$)   are the  Casimir 
  operator and  the third   generator  of  $sl_{H}(2)$ (resp.  
  $sl_{V}(2)$) for the trinucleotide state or codon,
\be 
 \hat{X}  \; \psi(XYZ) =  r(XYZ) \; \psi(XYZ) 
\ee
 where $\psi(XYZ) $ is the state $ \in V$, $V$ being the space of the 
 irreps. of $U_{q \to 0}(sl_{H}(2) \oplus sl_{V}(2))$, corresponding 
 to the XYZ codon and, using the 
 same notation for the operators and for their eigenvalues,
 \be
  r =  [  \, \alpha  \,   Q^{1}   \, - \, \beta \, 
  J_{3,V}^{1} (J_{3,V}^{1} - 1) \,
  +  \,  4  \gamma  \, (C_{H} + C_{V}) ] \; 2  \, (J_{3,H}  + \eta J_{3,V})  
\label{eq:fe} 
\ee
 the values of the quantities appearing in eq.(\ref{eq:fe}) are given 
 in Table \ref{table:dinuc} and Table \ref{table:rep}.
 \begin{table}[htbp]
 \small 
\centering
\caption{$\Big.$ Dinucleotides representation content and charge Q}
\label{table:dinuc}
\begin{tabular}{crrrrccrrrrc} 
\hline
dimer & $J_{H}$ & $J_{V}$ & $J_{3H}$ & $J_{3V}$ & $Q$ & dimer & $J_{H}$ & $J_{V}$ 
& $J_{3H}$ & $J_{3V}$ & $Q$ \\
\hline
CC & 1 & 1 & 1 & 1 & 7 & GC & 1 & 1 & 1 & 0 & 5\\
CU & 0 & 1 & 0 & 1 & 1 & GU & 0 & 1 & 0 & 0 & 1\\
CG & 1 & 0 & 1 & 0 & 3 & GG & 1 & 1 & 1 & $-1$ & 3\\
CA & 0 & 0 & 0 & 0 & $-1$ & GA & 0 & 1 & 0 & $-1$ & $-1$ \\
UC & 1 & 1 & 0 & 1 & 3 & AC & 1 & 1 & 0 & 0 & 1 \\
UU & 1 & 1 & $-1$ & 1 & $-1$ & AU & 1 & 1 & $-1$ & 0 & $-3$ \\
UG & 1 & 0 & 0 & 0 &$-1$ & AG & 1 & 1 & 0 & $-1$ & $-1$ \\
UA & 1 & 0 & $-1$ & 0 & $-5$ & AA & 1 & 1 & $-1$ & $-1$ & $-5$ \\
\hline
\end{tabular}
\end{table}
The transition matrix between the codon $i = XYZ$ and the codon $j = X'Y'Z'$
is
\be
 Q_{ji} = F(d_{ji}) \; q_{ji} \;\;\;\;\;\;\;\;\;\;\;\; j \neq i
 \label{eq:t}
\label{eq:trans}
\ee
where $F(d_{ji})$, the strength of the transition, is a {\bf decreasing} function  
 of the argument and $d_{ji}$ is the distance between the initial and final codon
\be
d_{ji} = | r(X'Y'Z')  \, - \,  r(XYZ)|
\label{eq:dis}
\ee
and $q_{ji}$ is  the  element of a matrix {\bf q} such that
\be
q_{ji} = 1  \;\;\; \mbox{i,j nearest codons} \;\;\;\;\;\; 
q_{ji} = 0  \;\;\; \mbox{otherwise}
\ee
   If the strength are considered as constants, our model is 
essentially equivalent to a reversible Markov model with constant parameters. 
A few words to justify the assumptions eq.(\ref{eq:fe}). Of course there 
is an arbitrary infinite way of defining the correspondence between a 
codon and a point of an Euclidean space. Our choice  is such that
 to a larger variation of the charge, i.e. to a larger variation of 
  the physico-chemical properties, corresponds a larger distance and 
   that the distance between codons in the same irrep.
is lower that codon in differents irreps.. 
  Generally, from eq.(\ref{eq:fe}), the distance between two codons, differing  by a nucleotide in the 
middle position or in the first position, is larger, due to the 
change of the value of the charge, than the distance between two codons, 
differing  by a nucleotide in the third position.
   At this stage our model can be considered as a markovian model with 
neighbors depending parameters.  
 
\section{Amino acid substitution matrices}
    
In this section we recall the definition and the differences between the 
experimentally determined mutation matrix.
The sequences alignment of proteins is a most powerful tool to get 
insights on the protein functions and to compute substitution rates due to
 evolutionary processes. The first
scheme was proposed in the  seventies by M. Dayhoff \cite{Dayhoff} and it is 
generally considered as the standard scheme. It is based on the alignments 
of protein sequences that are at least 85 \% identical. 
The evolutionary distance in measured in ``accepted point mutation" (PAM). 
Two sequences are said to be 1 PAM distant if they differ on average by one 
accepted-point mutation per 100 amino acids. The term ``accepted" means 
that the mutation of the amino acid has been incorporated into the 
protein's progeny, i.e. the mutation has not produced harmful consequences.
The original Dayhoff matrix, by construction,  was biased by the sample of
 proteins available at that time, mainly small globular preoteins, and 
  emphasized the rate of mutation in 
the  highly mutable amino acids. Another shortcoming of 
this scheme is that relationships between far distant sequences are poorly 
inferred, due to to the presence of deletions and insertions.
 A matrix, taking into account substitutions poorly 
 represented in the original Dayhoff's analysis and making use of a 
 statistics about 35 times higher, was computed in ref. 
 \cite{JTT} and it is known as PET91. 
\footnote{To study the relations for distant sequences a more reliable model has been 
proposed  in 1992 \cite{HH}, which is presently known as block substitution 
 matrix (BLOSUM).} 
 We make a comparison between our 
data and the 1-PAM PET91 matrix, see Table II of \cite{JTT}. 
In that table the data are referred to the substitution of the 
amino acids, so we cannot compare them directly with our predictions, 
which refer to the codon mutations. We have to consider for each amino acid the 
multiplet of codons encoding it and then to consider only the 
one-nucleotide mutations. In this process we have  to take into 
account the preferred codon usages, which depend on the biological species 
and on the type of gene analysed. In this preliminary analysis we make the simple 
(and definitely incorrect)
assumption of an uniform codon usage. The  experimental Dayhoff matrix 
entries between the amino acids $a$ and $b$ are identified as
 \be
 M_{ab} = \sum_{i,j} \, f_{i}^{a} \,  M_{ab}^{ij}
 \ee
where $f_{i}^{a}$ is the frequence of the $i$ codon in the amino 
acid $a$,$ M_{ab}^{ij}$ is the substitution rate matrix for the codons and 
the sum is over all the codons encoding the amino acids $a$ and 
$b$ differing by only one nucleotide. 
 The comparison with experimental data requires  one more
assumption.  We have to compare the matrix 
\be
{\bf P}(t) = \exp{{\bf Q} \, t}  \label{eq:Pt}
\ee
with the $x-PAM$-mutation matrix ${\bf M}_{x-PAM}$ which is computed at
a $x$ distance between the amino acids sequences. Commonly $1-PAM$ 
evolutionary distance is considered to correspond to a time interval of $ 
\approx 1 x 10^{7}$ years and the correspondence between the PAM matrix and the 
instantaneous rate matrix is 
\be
{\bf M}_{1-PAM} = \exp{{\bf Q} \, t} \;  \approx  \; {\bf 1} + 0.1 {\bf Q}
\ee  
i.e. the unit of time is choosen $ \tau_{0} = 1 x 10^{8}$ years.
It should be remarked that the above matrices, by construction are really 
divergence matrices, that is  they provide the probability that 
the $j$ state   in the first sequence,  will be replaced  by the $i$ state in the
 second xPAM  distant sequence. Moreover these matrices have been 
 build up assuming a symmetric probabily of mutation between two 
 amino acids and, consequently, the estimated rate is lower for the amino 
 acid which has a larger frequency. Therefore, strictly speaking, a 
 direct comparison between the  rate matrix eq.(\ref{eq:t}) and the 
 amino acid substitution matrices is uncorrect. However, as in the 
 present work we present only semiquantitative comparison, our 
 conclusions should not be sensibly affected by the above remarks.

\section{Predictions of model}

\subsection{Stability}

From the assignment of the codons to the different irreps., see 
 Table \ref{table:rep}, and the  assumed distance, see eq.(\ref{eq:dis}),  
 we can make a set of general predictions independent of the 
 structure of the $F$ function and of the detailed values of  
 $\alpha, \beta, \gamma$ and $\eta$. 
Considering a single-nucleotide mutation, each codon can make transition 
in the (9) nearest codons. Some of these codons can be synonimous (silent
mutations) or stop codons (nonsense mutations), both being unobservable in 
the framework of the substitution matrices. However, without a thorough 
analysis of their physico-chemical properties and/or their functional 
functions, we should expect   amino acids encoded by multiplets of the same 
dimension  to be approximately equally stable, i.e. the diagonal entries
of the mutation matrix {\bf M} should be of the same order. In the crystal 
basis model, see Table \ref{table:rep}, not all the codons are on the same foot as they 
belong to different  irreps. spaces. We indeed expect that  mutations between 
codons in the same  irrep. to occur more frequently than  mutations between 
different  irreps., provided that the 
values of  $ J_{3,V}^{1}$  are  close and the signs of their charge  
$Q$  are the same. This requires that we 
have to compare respectively long multiplets and short 
multiplets between them. Moreover in each fixed space, 
the codons represented 
by highest or lowest weight are ``surrounded" by a smaller number of nearest 
codons.  From an analysis of the  positions of the codons in 
the different irreps., we can qualitatively, from eq.(\ref{eq:fe}), derive a 
hierarchy in the stability.
\bea
& Gly > Pro > Ala > Thr > Ser* \nonumber \\
& Phe > LysI > le**  > Asn \nonumber Ê\\
& Leu* > Val \;\;\;\;\;\; Glu > Asp \nonumber \\
& His \approx Gln \;\;\;\;\;\; Trp >> Met
\eea 
where the * (**) is written to recall that we are dealing with a sextet 
(triplet), so our analysis  is less reliable.
A comparison with the experimental data from the PET91 and  Dayhoff 
matrices for the average mutability, see Table \ref{table:stab}, shows a remarkably satisfactory 
aggreement (higher stability implies lower mutability). Note that the comparison
 between His and Gln which, at first 
sight, is  not  satisfactory with the Dayhoff data, should be analyzed on the light 
 of   the wide 
range of variation  of the values of 
the average relative mutability  for the doublets (between  20 and 134).
 A more detailed analysis should 
require an evaluation of the form of the $F$ functions and of the values of 
the constants appearing in eq.(\ref{eq:fe}).

\begin{table}[htbp]
\centering
\caption{Relative mutabiliity for the 20 amino acids  
with respect to Ala, arbitrarily fixed to 100, from Table III of \cite{JTT}.}
\label{table:stab}
\footnotesize
 \begin{tabular}{crrrrc} 
\hline
 amino acid & $PET91$ & Dayhoff & amino acid & $PET91$ & Dayhoff \\
[1mm] 
\hline 
  Ala  &  100 &  100  & Leu  &  54 &  40  \\
 Arg  & 83 &  65 & Lys  & 72 &  56 \\
 Asn  &  104 & 134 & Met  &  93 &  94 \\
 Asp  &  86 &  106  & Phe &  51 &  41 \\ 
 Cys  & 44 &  20  & Pro  &  58 &  56    \\
 Gln  &  84 &  93   & Ser  &  117 &  120   \\
 Glu  &  77 &  102  & Thr  &  107 &  97   \\
 Gly  &  50 &  49 &  Trp  &  25 &  18    \\ 
 His  &  91 & 66 & Tyr  &  50 &  41  \\
 Ile  &  103 &  96 & Val  &  98 &  74    \\
 [1mm]\hline
\end{tabular}
 \end{table}

\subsection{Relation between rates}

In the following  we use the 
 standard notation Y = C, U (pyrimidines) and R = G, A (purines) and N 
 for any nucleotide.
 First we look for qualitative prediction for the rate of transition 
 between two amino acids $a$ and $b$ (R$(a \lra b)$) which follow directly 
 from 
 eq.(\ref{eq:t}) and from the assumed behaviour of the $F$ function,  without any 
 information of the values of $\alpha, \beta, \gamma$. 
 Fron an  inspection of  eqs.(\ref{eq:dis}), (\ref{eq:fe}), (\ref{eq:Q}) 
 and   Tables \ref{table:dinuc}, \ref{table:rep}, we  can write
 a set of inequalities between the rates for several amino acids.  
 The results of 
 our analysis are reported in 
 Table \ref{table:results} where for 
 any couple of amino acids we write the experimental values (Exp) taken from PET 
 matrix \cite{JTT}. 
 Of course we cannot make any more precise statement on the range of 
the inequalities, due to the yet undefined $F$ function.
 From the experimental data that $R(Phe \lra  Leu)  >  R(Phe \lra 
 Tyr)$ (Exp.: 230 | 179) we derive $ \eta > 2$. Then we expect
  \begin{eqnarray}
 R(Ala \lra  Pro) & <   & R(Ala \lra Val) \;\;\;\;\;\;\; \mbox{Exp: 
 ~ 23 | 193}   
  \end{eqnarray} 
  Let we remark that the following mutations between doublets: $ Asn \lra Lys$ 
 (AAY $ \lra $ AAR), $Asp \lra Glu$  (GAY $ \lra $ GAR), $His \lra Gln$
  (CAY $ \lra $ CAR), share the common features to involve a mutation 
  in the 3rd nucleotide and to have the same 2nd nucleotide A. So  from 
  the   assumption that the middle nucleotide is the one which 
  strongly determines the physico-chemical properties, 
  comparable mutation rates   should  be
  expected. On the contrary in our model, 
  from eq.(\ref{eq:fe}), we expect different rates, except for a 
  numerical coincidence for at most two of the considered mutations.
  The experimental rates are different (resp.: 150 | 478 | 233). So
  we derive the following inequality: 
  \begin{equation}
 |60 \gamma - 4 \beta - 10 \alpha|  > |12 \gamma - 2 \alpha| >
 |36 \gamma - 4 \beta - 2 \alpha|
  \label{eq:ineq}
  \end{equation}
  
Let us note that our analysis puts into evidence:

   \noindent a) a dissimilarity 
  between the transversions $ C \lra A$ and $ U \lra G$, which 
  apparently has not before either remarked;
  
  \noindent b) a "penalty", in the form of an increase of the distance, 
     appears for mutations   between codons  with
 $|J_{3,H}|$  or $|J_{3,V}|$ $>$ 1/2.
   
 Let us recall once more that in the determination of the mutation 
 rate the mutability, the frequency of occurrence and the codon 
 distribution frequency of the considered amino acid play a role.
   \section{Conclusions}
 It is believed that the mutations are essentially random effects, 
especially in the non coding seuqences. For the coding sequences it is 
known   the presence of evolutionary bias. Our analysis concerns only 
the coding sequences and  provides  indication of the presence of general 
 pattern and symmetry, not before observed. By trial and errors, 
 following the leading idea to incorporate in a   suitable 
 {\bf metric}  in a n-dim. space the effects of the near neighbours and the
 influence of the physico-chemical properties of the different amino acids in the
  rate mutation, we have build a simple model which is able to reproduce in a 
semi-quantitative way the hierarchy of the most frequently observed mutation between 
amino acids. The predictions  well agree with the experimental data of PET91. 
One should 
 check that    no inconsistency appears in the computed 
   inequalities. This is true for the reported set, but it has to be 
   carefully checked for all the mutations rates. It should also be noticed that the model is able to explain some 
puzzling features, for example:
\begin{enumerate} 
\item the almost equality of the rates  $R(Gly \lra Asp)$  and  
$R(Gly \lra Arg)$ (Exp.: 70), the first  mutation  resulting from the 
  transition of the 1st nucleotide, 
$GGR \lra AGR$, and the second from the  transitions of the 2rd nucleotide,
  $GGR \lra GAR$;
\item the fact that $R(Gln \lra His)$  ($CAR \lra CAY$, transversion of 
the 3rd nucleotide) is lower than 
$R(Gln \lra Glu)$  ($CAR \lra GAR$,  transversion of the 1st nucleotide)
\item  the fact that $R(Ser \lra  Thr)$ is lower than $R(Ser \lra Ala)$
  although any codon of the sextet $Ser$ can  go into the multiplet 
  encoding $Thr$ by 
  single nucleotide change while only the codons of the quartet UCN can  go 
  into the multiplet encoding $Ala$, by single nucleotide change. 
\end{enumerate} 
 A more quantitative analysis requires to take into 
account the normalisation of the transition matrix
\begin{equation}
\sum_{j} \, Q_{ji} = 1 \;\;\;\;\;\;\;\; \forall i
\end{equation}
and to evaluate the function $F$ of eq.(\ref{eq:fe}). Moreover one 
should know the codon usage frequency.
The parametrization in terms of only 4 parameters (which indeed can be 
reduced to 3 as one can be absorbed in the function $F$) and the 
identification of a codon with  a real number may
be a too simple choice. Going on with the analysis, likely, one will 
face   some inconsistencies between the 
theoretical relations.   Hopefully these pathologies can be cured 
with slight modifications of    eqs.(\ref{eq:fe}) 
and (\ref{eq:dis}).  

It is appropriate to underline that  
 this approach can be easily generalized to describe more complex phenomena, neglected in 
this paper, as the multiple nucleotide changes, the observed presence of hotspots 
for the mutations, the variation of the mutations with the type of 
proteins, the probable occurrence of spurts in the evolution,
the scaling behavior of the mean parameter substitution in 
function of the total length of genome \cite{NS}, etc.
 A criticism  can be raised against this model:   it is
 essentially  based on the properties of the genetic code while the 
accepted mutations are the  replacement of an amino acid by a similar one.
Some of the chemical properties which mostly influence the chances 
of mutations, like the hydrophobicity, charge, size, are related to the 
genetic code, \cite{FSSp}, but   
many of the physical chemical properties of the amino acids are believed to 
have been  more imposed by natural selection than by genetic code constraints.
If the plausibility of the model is confirmed, this arises a puzzling 
question. The comparison for the mutation rates  between the predicted 
values of the theoretical time evolution operator {\bf P(t)} and the 
experimental values of the evolution distance matrix {\bf M}, which can be 
criticized from many points of view, has been done as the amino acid mutation 
matrix is, at my knowledge, the only source of mutation data with a 
large statistics, obtained by analysing many thousands of proteins.

 \newpage
 \begin{table}[htbp]
\caption{The eukaryotic or standard code code. Upper labels denote 
different irreps.}
\label{table:rep}
\footnotesize
\begin{center}
\begin{tabular}{ccrrrrccrrrr} 
\hline 
codon & amino acid & $J_{H}$ & $J_{V}$ & $J_{3,H}$ & $J_{3,V}$& codon & 
amino acid & $J_{H}$ & $J_{V}$ & $J_{H,3}$ & $J_{V,3}$ \\
\hline
CCC & Pro P & $3/2$ & $3/2$ & $3/2$ & $3/2$ & UCC & Ser S & $3/2$ & $3/2$ & 
$1/2$ & $3/2$ \\
CCU & Pro P & $(1/2$ & $3/2)^1$ & $1/2$ & $3/2$ & UCU & Ser S & $(1/2$ & 
$3/2)^1$ & $-1/2$ & $3/2$ \\
CCG & Pro P & $(3/2$ & $1/2)^1$ & $3/2$ & $1/2$ & UCG & Ser S & $(3/2$ & 
$1/2)^1$ & $1/2$ & $1/2$ \\
CCA & Pro P & $(1/2$ & $1/2)^1$ & $1/2$ & $1/2$ & UCA & Ser S & $(1/2$ & 
$1/2)^1$ & $-1/2$ & $1/2$ \\[1mm] \hline
CUC & Leu L & $(1/2$ & $3/2)^2$ & $1/2$ & $3/2$ & UUC & Phe F & $3/2$ & 
$3/2$ & $-1/2$ & $3/2$ \\
CUU & Leu L & $(1/2$ & $3/2)^2$ & $-1/2$ & $3/2$ & UUU & Phe F & $3/2$ & 
$3/2$ & $-3/2$ & $3/2$ \\
CUG & Leu L & $(1/2$ & $1/2)^3$ & $1/2$ & $1/2$ & UUG & Leu L & $(3/2$ & 
$1/2)^1$ & $-1/2$ & $1/2$ \\
CUA & Leu L & $(1/2$ & $1/2)^3$ & $-1/2$ & $1/2$ & UUA & Leu L & $(3/2$ & 
$1/2)^1$ & $-3/2$ & $1/2$ \\[1mm] \hline
CGC & Arg R & $(3/2$ & $1/2)^2$ & $3/2$ & $1/2$ & UGC & Cys C & $(3/2$ & 
$1/2)^2$ & $1/2$ & $1/2$ \\
CGU & Arg R & $(1/2$ & $1/2)^2$ & $1/2$ & $1/2$ & UGU & Cys C & $(1/2$ & 
$1/2)^2$ & $-1/2$ & $1/2$ \\
CGG & Arg R & $(3/2$ & $1/2)^2$ & $3/2$ & $-1/2$ & UGG & Trp W & $(3/2$ & 
$1/2)^2$ & $1/2$ & $-1/2$ \\
CGA & Arg R & $(1/2$ & $1/2)^2$ & $1/2$ & $-1/2$ & UGA & Ter & $(1/2$ & 
$1/2)^2$ & $-1/2$ & $-1/2$ \\[1mm] \hline
CAC & His H & $(1/2$ & $1/2)^4$ & $1/2$ & $1/2$ & UAC & Tyr Y & $(3/2$ & 
$1/2)^2$ & $-1/2$ & $1/2$ \\
CAU & His H & $(1/2$ & $1/2)^4$ & $-1/2$ & $1/2$ & UAU & Tyr Y & $(3/2$ & 
$1/2)^2$ & $-3/2$ & $1/2$ \\
CAG & Gln Q & $(1/2$ & $1/2)^4$ & $1/2$ & $-1/2$ & UAG & Ter & $(3/2$ & 
$1/2)^2$ & $-1/2$ & $-1/2$ \\
CAA & Gln Q & $(1/2$ & $1/2)^4$ & $-1/2$ & $-1/2$ & UAA & Ter & $(3/2$ & 
$1/2)^2$ & $-3/2$ & $-1/2$ \\[1mm] \hline
GCC & Ala A & $3/2$ & $3/2$ & $3/2$ & $1/2$ & ACC & Thr T & $3/2$ & $3/2$ & 
$1/2$ & $1/2$ \\
GCU & Ala A & $(1/2$ & $3/2)^1$ & $1/2$ & $1/2$ & ACU & Thr T & $(1/2$ & 
$3/2)^1$ & $-1/2$ & $1/2$ \\
GCG & Ala A & $(3/2$ & $1/2)^1$ & $3/2$ & $-1/2$ & ACG & Thr T & $(3/2$ & 
$1/2)^1$ & $1/2$ & $-1/2$ \\
GCA & Ala A & $(1/2$ & $1/2)^1$ & $1/2$ & $-1/2$ & ACA & Thr T & $(1/2$ & 
$1/2)^1$ & $-1/2$ & $-1/2$ \\[1mm] \hline
GUC & Val V & $(1/2$ & $3/2)^2$ & $1/2$ & $1/2$ & AUC & Ile I & $3/2$ & 
$3/2$ & $-1/2$ & $1/2$ \\
GUU & Val V & $(1/2$ & $3/2)^2$ & $-1/2$ & $1/2$ & AUU & Ile I & $3/2$ & 
$3/2$ & $-3/2$ & $1/2$ \\
GUG & Val V & $(1/2$ & $1/2)^3$ & $1/2$ & $-1/2$ & AUG & Met M & $(3/2$ & 
$1/2)^1$ & $-1/2$ & $-1/2$ \\
GUA & Val V & $(1/2$ & $1/2)^3$ & $-1/2$ & $-1/2$ & AUA & Ile I & $(3/2$ & 
$1/2)^1$ & $-3/2$ & $-1/2$ \\[1mm] \hline
GGC & Gly G & $3/2$ & $3/2$ & $3/2$ & $-1/2$ & AGC & Ser S & $3/2$ & $3/2$ 
& $1/2$ & $-1/2$ \\
GGU & Gly G & $(1/2$ & $3/2)^1$ & $1/2$ & $-1/2$ & AGU & Ser S & $(1/2$ & 
$3/2)^1$ & $-1/2$ & $-1/2$ \\
GGG & Gly G & $3/2$ & $3/2$ & $3/2$ & $-3/2$ & AGG & Arg R & $3/2$ & $3/2$ 
& $1/2$ & $-3/2$ \\
GGA & Gly G & $(1/2$ & $3/2)^1$ & $1/2$ & $-3/2$ & AGA & Arg R & $(1/2$ & 
$3/2)^1$ & $-1/2$ & $-3/2$ \\[1mm] \hline
GAC & Asp D & $(1/2$ & $3/2)^2$ & $1/2$ & $-1/2$ & AAC & Asn N & $3/2$ & 
$3/2$ & $-1/2$ & $-1/2$ \\
GAU & Asp D & $(1/2$ & $3/2)^2$ & $-1/2$ & $-1/2$ & AAU & Asn N & $3/2$ & 
$3/2$ & $-3/2$ & $-1/2$ \\
GAG & Glu E & $(1/2$ & $3/2)^2$ & $1/2$ & $-3/2$ & AAG & Lys K & $3/2$ & 
$3/2$ & $-1/2$ & $-3/2$ \\
GAA & Glu E & $(1/2$ & $3/2)^2$ & $-1/2$ & $-3/2$ & AAA & Lys K & $3/2$ & 
$3/2$ & $-3/2$ & $-3/2$ \\[1mm] \hline
\end{tabular}
\end{center}
\end{table}

\newpage

\begin{table}[hb]
\centering
\caption[]{Theoretical inequalities for the rate mutations between two 
couples of amino acids. In the last two columns the experimental rate,
 from \cite{JTT},   for each couple.}
\label{table:results}
\footnotesize
\begin{tabular}{c|cc}
\hline
Theor: \,\, Rate(I) $<$ Rate(II)   & Exp-I & Exp-II \\
\hline
 $ R(Asp \lra Ala) $ $~ <$   $ R(Glu \lra Ala)$  &
   ~ 63  & ~Ê82 \\  
$  R(His \lra Pro)$ $ ~ < $  R$(Gln \lra Pro$) &
   ~ 58 & ~Ê81  \\  
$ R(Gly \lra Arg) $ $<$  $R(Gly \lra Ser)$   &
   ~ 70 & 129 \\ \ 
$ R(Gly \lra Asp) $ $< \approx$ $ R(Gly \lra Glu) $ 
& ~ 66 &~Ê70 \\  
$R(Trp \lra Arg)$  $ < \approx$ $ R(Met \lra Thr)$  
 & ~~ 7 & Ê123 \\ \ 
 $R(Gly \lra Arg)$  $ < \approx$ $ R(Gly \lra Glu)$  
 & ~ 70 & ~Ê70 \\  
$ R(Gln \lra Arg)$ $ <  $ $R(His \lra Arg)$ 
 & 154 & 164 \\  
$ R(Asn \lra Asp)$ $ <  $ $R(Asn \lra Ser)$   
&  284 & 344 \\  
$ R(Lys \lra  Gln)$ $ <  $ $R(Asn \lra His)$  
&  122 & 150 \\  
$ R(Lys \lra  Arg)$ $ <  $ $R(Asn \lra Ser)$  
&   334 & 344 \\ \ 
 $R(Ala \lra  Thr)$ $ <  $ $R(Ala \lra Ser)$   
 &  267 & 284 \\  
 $R(Met \lra  Thr)$ $ <  $ $R(Met \lra Val) $ 
 &  123 & 201  \\ \ 
 $R(Tyr \lra  Asp)$ $ <  $ $R(Tyr \lra Ser)$   
 & ~ 23 & ~ 43  \\  
 $R(Tyr \lra  Ser)$ $ <  $ $R(Tyr \lra His)$ 
 & ~ 43 & 134  \\  
 $R(Val \lra  Leu)$ $ <  $ $R(Val \lra Ala)$  
   & 161 & 226  \\   
$ R(Val \lra  Ala)$ $ <  $ $R(Val \lra Ile)$  
  & 226 & 504 \\  
$ R(Ser \lra  Thr)$ $ < $ $ R(Ser \lra Ala)$   
 &  278 & 297  \\  
$ R(Pro \lra  Thr)$ $ <  $ $R(Pro \lra Leu) $ 
  & ~ 69 &~ 97 \\   
$ R(Pro \lra  Thr)$ $ <  $ $R(Ser \lra Ala) $ 
  & ~ 69 & 297 \\  
  $ R(Pro \lra  Ala)$ $ <  $ $R(His \lra Arg) $ 
  & 150 & 164 \\  
   $ R(Ile \lra  Thr)$ $ <  $ $R(His \lra Arg) $ 
  & 149 & 164 \\  
   $R(His \lra Arg) $  $ <  $ $R(Pro \lra Ser) $ 
  & 164 & 190 \\   
  $ R(Thr \lra  Ile)$ $ <  $ $R(Thr \lra Ser) $ 
  & 134 & 325 \\  
\hline
\end{tabular}
 \end{table}

\end{document}